\documentclass[journal=nalefd,manuscript=letter]{achemso}

\usepackage[version=3]{mhchem} 
\usepackage{graphicx}
\usepackage{amssymb}
\usepackage{color}
\usepackage{amsmath}
\usepackage{bm}
\usepackage[breaklinks]{hyperref}
\usepackage[all]{hypcap}

\hypersetup{
    plainpages=false,
    bookmarks=false,         
    unicode=false,          
    pdfborder={0 0 0}
    pdftoolbar=true,        
    pdfmenubar=true,        
    pdffitwindow=false,     
    pdfstartview={FitH},    
    pdftitle={},    
    pdfauthor={},     
    pdfproducer={LNCMI-CNRS-UJF-UPS-INSA}, 
    pdfkeywords={High} {Magnetic} {Fields}, 
    pdfnewwindow=true,      
    linktoc=section,
    colorlinks=true,       
    linkcolor=blue,          
    citecolor=red,        
    filecolor=magenta,      
    urlcolor=blue           
}

\author{J. Jadczak}
\affiliation{Laboratoire National des Champs Magn\'etiques Intenses, CNRS-UJF-UPS-INSA, 143, avenue de Rangueil, 31400
Toulouse}\altaffiliation{Institute of Physics, Wroclaw University of Technology, 50-370 Wroclaw, Poland}

\author{P. Plochocka}
 \email{paulina.plochocka@lncmi.cnrs.fr}\affiliation{Laboratoire National des Champs Magn\'etiques Intenses,
CNRS-UJF-UPS-INSA, 143, avenue de Rangueil, 31400 Toulouse}

\author{A. Mitioglu}
\affiliation{Laboratoire National des Champs Magn\'etiques Intenses, CNRS-UJF-UPS-INSA, 143, avenue de Rangueil, 31400
Toulouse} \affiliation{Institute of Applied Physics, Academiei Str. 5, Chisinau, MD-2028, Republic of Moldova}

\author{I. Breslavetz}
 \affiliation{Laboratoire National des Champs Magn\'etiques Intenses,
CNRS-UJF-UPS-INSA, 25, avenue de Martyrs, 38042 Grenoble}

\author{M. Royo}
 \affiliation{Institute for Nanoscience, CNR-NANO S3, via Campi 213/A, 41125 Modena, Italy}
 \affiliation{Departament de Qu\'imica F\'isica I Anal\'itica, Universitat Jaume I, Av. Sos Baynat s/n, 12080 Castell\'o, Spain}

\author{A. Bertoni}
 \affiliation{Institute for Nanoscience, CNR-NANO S3, via Campi 213/A, 41125 Modena, Italy}

\author{G. Goldoni}
 \affiliation{Institute for Nanoscience, CNR-NANO S3, via Campi 213/A, 41125 Modena, Italy}
 \affiliation{Department of Physics, Informatics and Mathematics, University of Modena and Reggio Emilia, Italy}

\author{T. Smolenski}
\affiliation{Institute of Experimental Physics, Faculty of Physics, University of Warsaw, Ho\.za 69, 00-681 Warsaw,
Poland}

\author{P. Kossacki}
\affiliation{Institute of Experimental Physics, Faculty of Physics, University of Warsaw, Ho\.za 69, 00-681 Warsaw,
Poland}

\author{A. Kretinin}
\affiliation{Department of Condensed Matter Physics, The Weizmann Institute of Science, Rehovot 76100, Israel}
\affiliation{School of Physics and Astronomy, University of Manchester, UK}

\author{Hadas Shtrikman}
\affiliation{Department of Condensed Matter Physics, The Weizmann Institute of Science, Rehovot 76100, Israel}

\author{D. K.  Maude} \affiliation{Laboratoire
National des Champs Magn\'etiques Intenses, CNRS-UJF-UPS-INSA, 143, avenue de Rangueil, 31400 Toulouse}

\title[\texttt{achemso} demonstration]
{Unintentional high density p-type modulation doping of a GaAs/AlAs core-multi-shell nanowire}

\begin{document}
\begin{abstract}
Achieving significant doping in GaAs/AlAs core/shell nanowires (NWs) is of considerable technological importance but
remains a challenge due to the amphoteric behavior of the dopant atoms. Here we show that placing a narrow GaAs quantum
well in the AlAs shell effectively getters residual carbon acceptors leading to an \emph{unintentional} p-type doping.
Magneto-optical studies of such a GaAs/AlAs core multi-shell NW reveal quantum confined emission. Theoretical
calculations of NW electronic structure confirm quantum confinement of carriers at the core/shell interface due to the
presence of ionized carbon acceptors in the 1~nm GaAs layer in the shell. Micro-photoluminescence in high magnetic
field shows a clear signature of avoided crossings of the $n=0$ Landau level emission line with the $n=2$ Landau level
TO phonon replica. The coupling is caused by the resonant hole-phonon interaction, which points to a large 2D hole
density in the structure.
\end{abstract}

 \vspace{0.5cm}


Semiconductor nanowires (NWs) represent a rapidly expanding field of research largely due to their great technological
promise~\cite{Hu99,Cui01,Therlander06,Lieber07,Lu07}. For example, transistor action has been demonstrated using carbon
nanotubes and silicon nanowires,~\cite{Tans98, Cui03} it has been suggested that indium phosphide nanowires can be used
as building blocks in nanoscale electronics,~\cite{Duan01} and doped radial core-multi-shell NWs have good chances to
find industrial applications as high efficiency solar cells~\cite{Krogstrup13, Spirkoska11,Fickenscher13,Estrin13,
FontcubertaiMorral08}. NWs with two dimensional (2D) carriers localized at the core/shell interface offer new
perspectives in quantum electronics.~\cite{Lieber07} However, to introduce carriers the control and understanding of
the doping mechanisms in NWs is crucial.

Epitaxial GaAs has been investigated for more than 40 years, a field of research propelled by the discovery of the
quantum Hall effect with its panoply of exotic many body ground states.~\cite{Klitzing1980,Tsui1982} Significant
efforts have been made to obtain high mobility 2D carriers in GaAs heterostructures grown by molecular beam epitaxy
(MBE). Notably, the development of remote or modulation doping, which spatially separates the carriers from the dopant
atoms, was a crucial step in the discovery of the fractional quantum Hall effect.~\cite{Tsui1982} The direct
application of the modulation doping techniques to GaAs/AlAs NWs would seem to be a natural evolution. Today high
quality GaAs/AlAs nanowires with a large aspect ratio and typical diameters of a few tens of nanometers are routinely
grown by MBE using the vapour-liquid-solid (VLS)
method\cite{Titova06,Pistol08,Shtrikman09a,Shtrikman09,Spirkoska09,Algra11,Jahn12,Musin12,Assali13}. While core multi
shell NWs would seem to be ideally suited for modulation doping, in practice achieving a significant doping remains a
challenge. In the VLS method, depending upon the growth plane, dopants can act as donors or acceptors~\cite{Hilse10}
and the dopant incorporation can be different for axial and lateral (sidewall) growth~\cite{Dufouleur10} resulting in
an inhomogeneous dopant distribution, or even compensation and negligible doping.~\cite{Casadei13}. In NWs the use of
the AlGaAs ternary alloy for the shell can lead to the segregation of the Ga and Al atoms leading to the formation of
the quantum dots.~\cite{Heiss13, Funk13} Moreover, using AlGaAs for the shell leads to a red shift of the
photoluminescence emission from the NW core, which is not understood.~\cite{Hocevar12} On the other hand both
experiment and theory suggest that modulation doping in core-multi-shell NWs can lead to non uniform charge
distribution with an accumulation of charge at the facets or corners of the hexagonal NW which can lead to quantum
confinement.~\cite{Funk2013}

In this work, we show that incorporating a narrow GaAs quantum well, which can be used to accommodate dopants, in the
AlAs shell of the NW can lead to significant unintentional p-type doping due to the incorporation of residual carbon
acceptors. Experiment and calculations demonstrate that charge transfer at low temperature leads to quantum confinement
with the formation of a \emph{high density} two dimensional hole gas at the core/shell interface of the NW.

\begin{figure}[h]
\includegraphics[width=8.5cm]{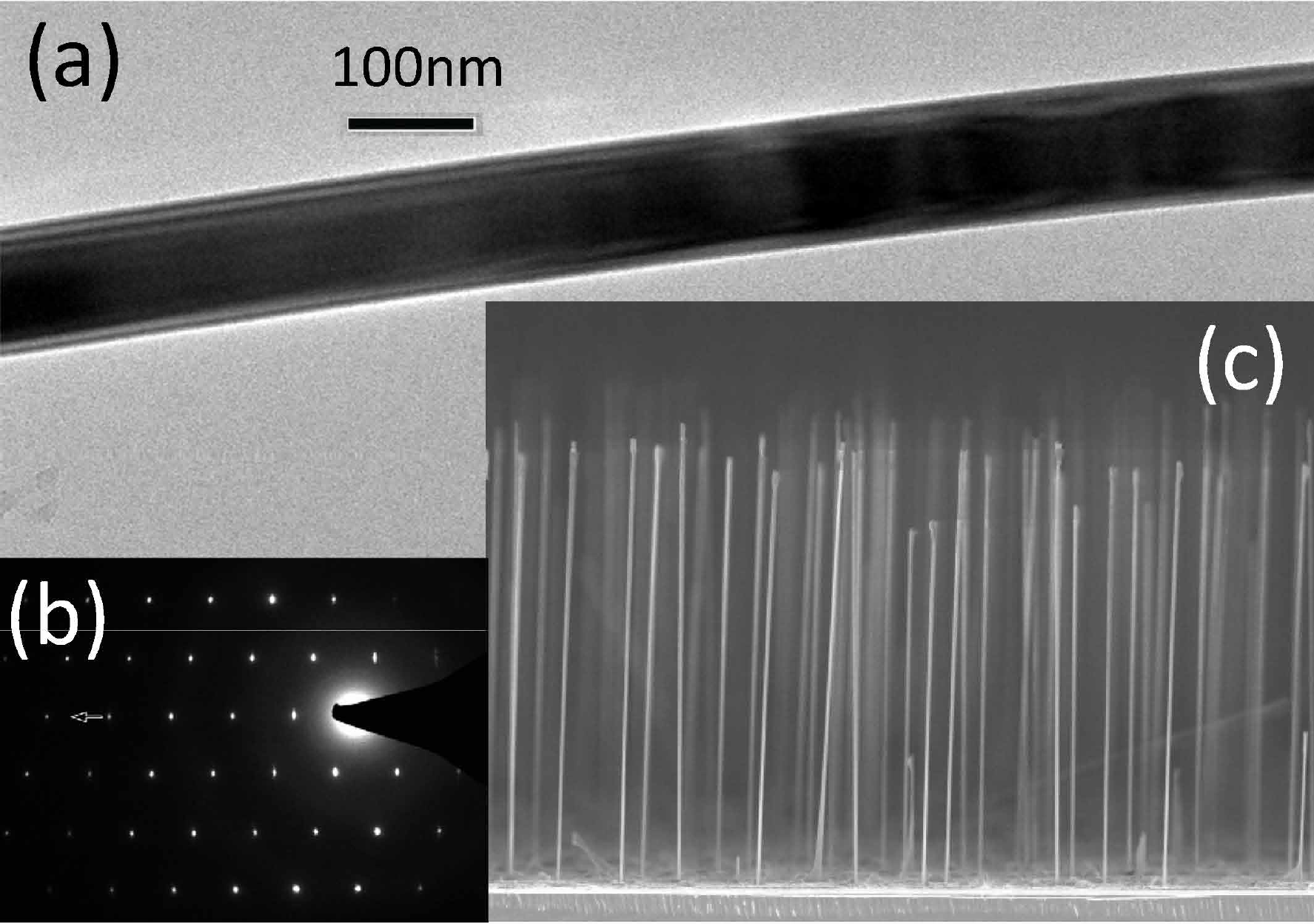}
\caption{\label{fig1} (a) typical transmission electron microscopy
(TEM) image taken from the center of a GaAs/AlAs core-multi-shell
NW, (b)pure zinc blende structure confirmed by the electron
diffraction taken from the [011] direction, the arrow is aligned
along the [111] growth direction, (c) scanning electron microscope
(SEM) image of the GaAs/AlAs core-multi-shell NWs $\sim 11 \mu$m
long.}
\end{figure}

The GaAs/AlAs NWs were grown by MBE using the self-assisted VLS method, on $(111)$-oriented silicon bearing a native
oxide layer. After water removal at~$200^{\circ}$C, the Si wafer was outgassed in a separate chamber ($600^{\circ}$C),
before being transferred into the MBE growth chamber. Growth was initiated by Ga condensation at pin holes in the
SiO$_{2}$ layer and carried out at~$640^{\circ}$C and a group V/III (As$_{4}$/Ga) ratio of $100$. Uniform diameter GaAs
NWs were grown with a high aspect ratio, no significant tapering and pure zinc-blende structure, as revealed by careful
transmission and scanning electron microscopy (Fig.~\ref{fig1}(a) and (c)) and electron diffraction taken from the
[011] direction (Fig.~\ref{fig1}(b)). For growth of the shell (nominally AlAs~3~nm/GaAs~1~nm/ AlAs~3~nm) and 12~nm GaAs
capping layer, the temperature was lowered to $520^{\circ}$C. The hexagonal GaAs NW core of side length $d \simeq
22~nm$ composes a substrate for the multilayer structure, which consists of a AlAs/GaAs/AlAs shell of a nominal
thickness of $7~nm$ and a GaAs capping layer. \textcolor{red}{}

\begin{figure}[h]
\includegraphics[width=8.5cm]{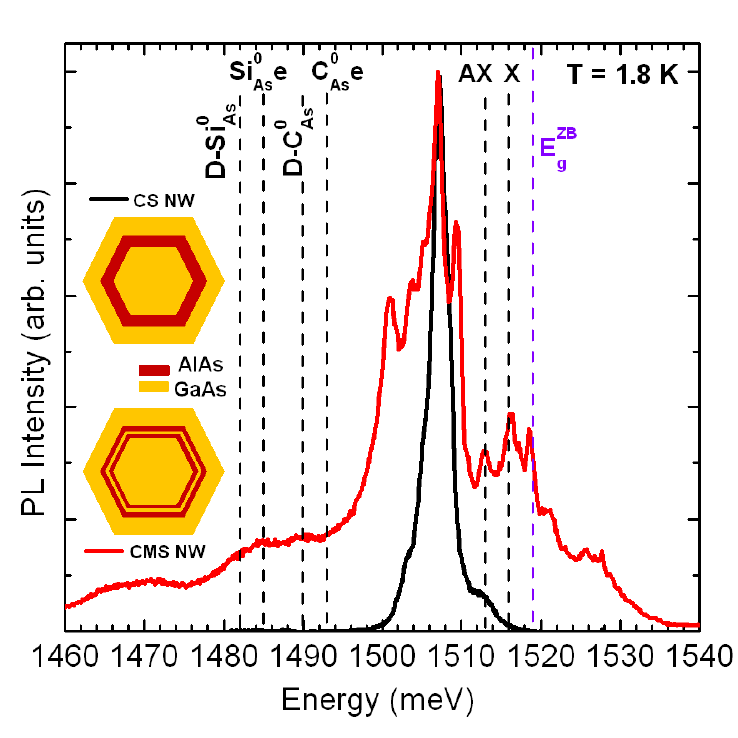}
\caption{\label{fig2} Typical $\mu$PL spectra of core-shell GaAs/AlAs NW (black line) and core-multi-shell GaAs/AlAs
(red line) excited by 660 nm laser line. Inset: schematic cross-section of core-shell and core-multi-shell NWs.}
\end{figure}

A typical $\mu$PL spectrum of a single core-multi-shell NW obtained in the absence of magnetic field at a temperature
$T=1.8$~K is presented in Fig.~\ref{fig2} (see methods section for experimental details). For comparison, we show a
$\mu$PL spectrum measured on a single core-shell NW.~\cite{Plochocka13} The structure of each sample is shown
schematically in the inset of Fig.~\ref{fig2}. We stress that both samples were grown in the same MBE chamber under
nominally the same growth conditions. Both NWs have a rather large core (80~nm for core-shell and 60~nm for
core-multi-shell) so that in a first approximation quantum confinement can be neglected. The common feature of both
emission spectra is the prominent peak at $1.507$~eV with shoulders which develop into well resolved emission lines in
a magnetic field.~\cite{Plochocka13} This transition is related to the so called KP series of excitons bound to defect
pairs with different separations in extremely high quality epitaxial GaAs.~\cite{Kunzel80} The emission energy of
$1.507$~eV corresponds well to the strongest line in the KP series in $1.504-1.511$~eV spectral
region.~\cite{Skolnick85}

The PL spectrum of the core-multi-shell NW is far richer than for the core-shell NW. In addition to the KP series, a
large number of emission lines are resolved already in the zero magnetic field spectrum. At higher energies but still
below the band gap of bulk GaAs ($E^{ZB}_{g}=1.519$~eV) we can distinguish the free exciton (X) at $1.515$~eV and
exciton bound to neutral acceptor (AX) at $1.513$~eV. Below the KP series, in the energy range $1.485-1.493$~eV we
observe a number of weak features with emission energies characteristic of the free-electron carbon-acceptor $A^{0}e$
and the donor carbon-acceptor ($D^{0}-A^{0}$) transitions.~\cite{Pavesi94} This suggests that we have doped regions
within the NW due to the unintentional incorporation of carbon. The shape of the spectra at low energy is
characteristic for p-doped structures, as recently observed for a single GaAs NW with an axial
heterojunction.~\cite{Sager13} In particular, it was shown that the emission energy for n-type and p-type material is
quite different; for p-type GaAs NWs the emission is dominated by recombination via acceptor centers, whereas for
n-type NWs emission is blue shifted with respect to the GaAs band gap due to a band filling effect with increasing
doping concentration.~\cite{Sager13, Borghs89}

For the core-multi-shell NWs we observe additional emission lines at energies higher than the band gap of GaAs in the
range $1.520-1.528$~eV. This emission energy is typical for a GaAs quantum well (QW) of width
$15-20$~nm~\cite{Ferreira1996,Glasberg2001} suggesting a quantum confinement of the carriers in the core. To
distinguish this high energy emission from the emission in other spectral regions we refer to it a ``2D like'' in the
rest of the paper. The core-multi-shell structure incorporates a 1~nm GaAs narrow quantum well in the AlAs shell at a
distance of 3~nm from the core. No emission from this ultra thin quantum well (QW) is detected at higher energies,
presumably due to the rapid thermalization (escape) of photo-created carriers. The 1~nm GaAs layer between the two AlAs
layers is expected to getter impurities, notably residual carbon atoms.~\cite{Meynadier85,Petroff84} It acts as an
efficient impurity trap due to the higher solubility of carbon atoms in GaAs and due to the floating of carbon atoms at
the AlAs vacuum interface during the MBE growth.~\cite{Meynadier85, Petroff84} Moreover, for AlAs we expect the
concentration of carbon to be quite high because Al atoms are more reactive with carbon or other
impurities.~\cite{Tokumitsu86} For example, the residual carbon incorporation in AlAs can be two orders of magnitude
higher than for GaAs.~\cite{Tokumitsu86} In our core-multi-shell NWs the carbon incorporated in the GaAs quantum well
can lead to a non-uniform charge distribution with excess holes accumulated at the core/shell interface.

\begin{figure}[h]
\includegraphics[width=12cm]{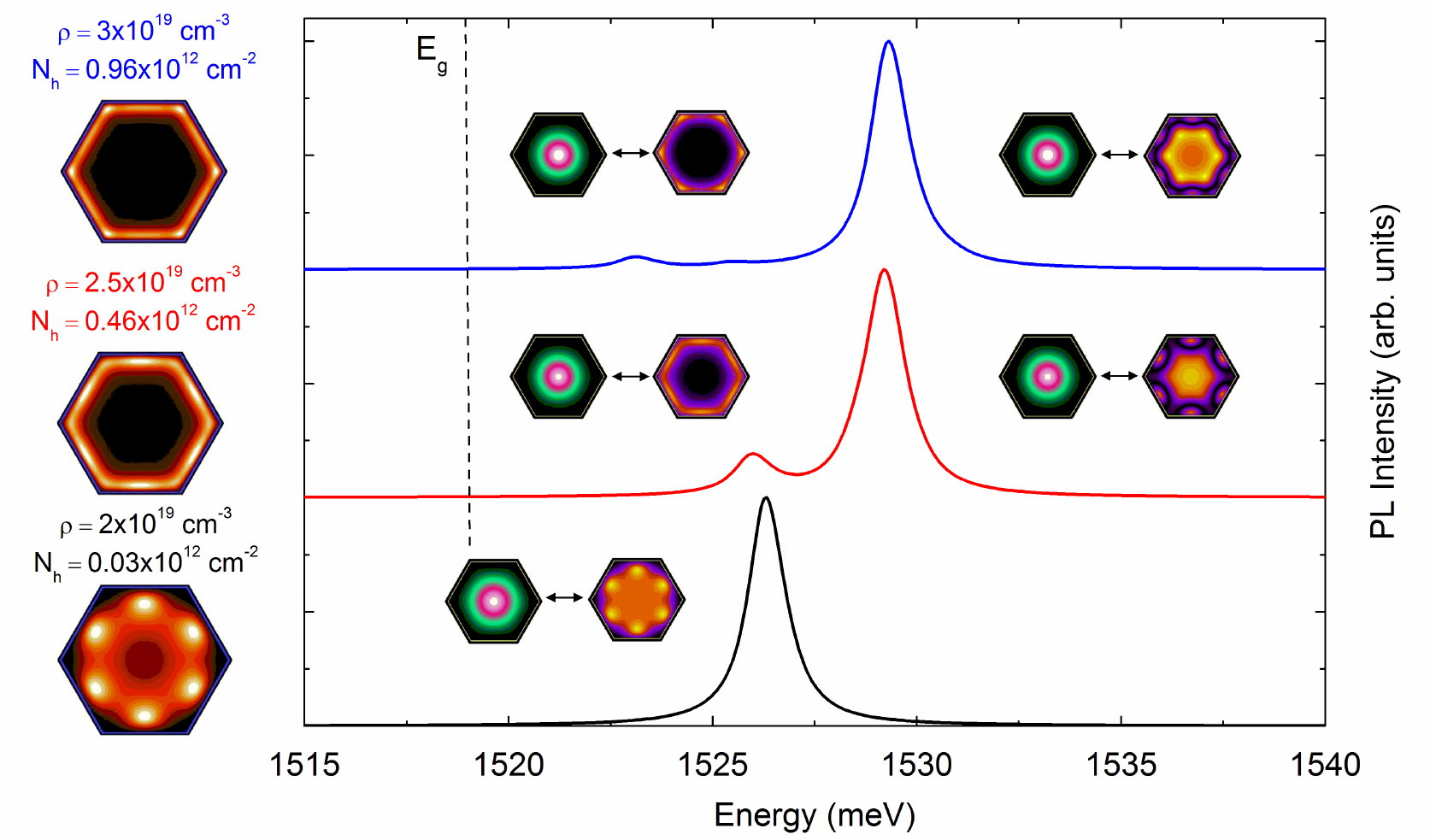}
\caption{Calculated PL spectra (main panel) and hole gas distributions (small left panels) for core-multi-shell NWs
with the p-doping densities $\rho_A$ indicated in the left panels, together with the 2D free hole density $N_h$. The
colormap insets illustrate the squared envelope functions for the electron (left) and hole (right) states whose
recombination yields the PL peaks. The vertical dashed line indicates the position of the GaAs bulk band gap.}
\label{fig3}
\end{figure}

This hypothesis is confirmed by 3D self-consistent simulations of the NW electronic structure. The modeling assumes
spatial invariance along the NW axis and includes the nominal multi-shell material modulation over the NW cross
section. The interstitial GaAs quantum well is uniformly doped with a constant density of acceptors $\delta_A$. Our
results predict that a hole gas starts to accumulate at the core/shell interface for p-doping densities of the order of
$2.0\times10^{19}$~$cm^{-3}$ (corresponding to a linear density of $4.63\times10^{7}$~$cm^{-1}$ in the NW). The hole
gas density distribution for a NW with three selected doping densities is illustrated in the left panels of
Fig.~\ref{fig3}. The different localization patterns are in line with previous results.~\cite{Bertoni11} At low
densities the hole gas is distributed near the heterojunction forming six wide channels at the center of the hexagonal
facets (bottom panel). As the doping density is increased the hole gas moves closer to the heterojunction and it forms,
first, a quasi-uniform sixfold bent 2DEG (middle panel), and finally, six tunnel-coupled narrow channels at the edges
(top panel).

The calculated PL spectra corresponding to the same doping densities can be seen in the main panel of Fig.~\ref{fig3}.
With increasing doping densities the exciton ground state red shifts and its intensity is substantially reduced. This
can be explained as follows: The peak red shifts due to the larger attracting mean field experienced by the electron
when the hole density is increased. At the same time, whereas the hole ground state is further localized near the
heterojunction, the electron ground state remains in the center of the GaAs core (see two top-left insets in
Fig.~\ref{fig3}) due to the repulsive potential generated by the high density of acceptors. Therefore, the overlap
between the ground electron and hole states becomes very small and the PL intensity of such transition is reduced. The
PL spectra of the NWs with higher acceptor densities show an additional intense peak at $1.529$~eV. As illustrated in
the corresponding insets in Fig.~\ref{fig3}, this peak originates from the recombination of an electron in the ground
state and a hole in an excited, but still occupied, state with an azimuthal-like nodal surface. Since this hole state
is well spread over the center of the GaAs core it has a large overlap with the electron ground state leading to an
intense peak in the PL spectrum. This peak is not observed at lower doping density since the excited hole state is
unoccupied.

\begin{figure}[h]
\includegraphics[width=8.5cm]{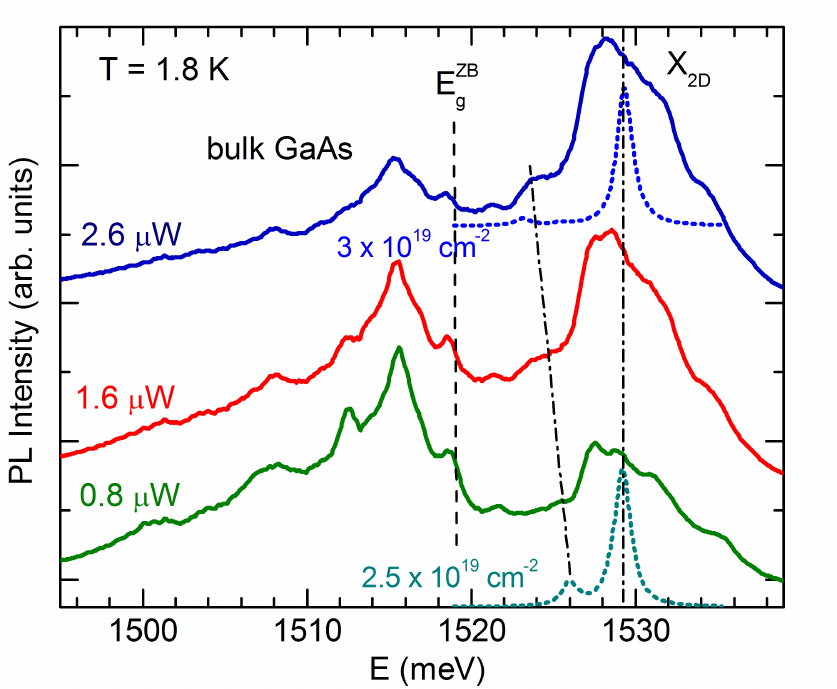}
\caption{\label{fig4}Evolution of the PL spectra (solid curves) of core-multi-shell GaAs/AlAs NW as function of the
excitation power, at T=~1.8~K. The calculated PL spectra for different doping levels are plotted for comparison (dashed
curves). The black dot-dashed lines are a guide to the eye to highlight the evolution of the position of 2D emission
lines described in the text. Note that the spectra have been shifted vertically for clarity.}
\end{figure}

The concentration of photo created electron hole pairs, and thus the concentration of holes can also be tuned by
varying the excitation power. A comparison between experimental results and theoretical simulation is presented in
Fig.~\ref{fig4}, which shows plots of $\mu$PL spectra for different excitation powers. The relative intensity of the KP
series of lines, below $E^{ZB}_{g}=1.519$~eV, decreases as the excitation power increases. This is due to the
saturation of these transitions when all the defect pairs have a bound exciton. In contrast, the 2D like emission
intensity increases continuously with excitation power. For the highest excitation power, the spectra are dominated by
emission related to the 2D hole gas. This further confirms the very different origins of each emission channel.

Theory predicts that the mechanism of the 2D recombination depends on the concentration of photo-excited carriers.
Increasing the number of photo created electron-hole pairs changes the overlap between electron and ground hole states
due to the modified distribution of 2D hole gas in the core. The calculated PL spectra for two different hole
concentrations are shown in Fig.~\ref{fig4} for comparison, showing qualitative agreement with the experimental data.
The transition around 1.526 eV shifts toward lower energies with increasing excitation power, while the second peak
around $1.529$~eV, originated from recombination of electrons with holes in excited state, remains at the same energy.
Its intensity is significantly enhanced as a straightforward consequence of the increased occupation of the electron
and hole sub-bands involved in the radiative recombination.

\begin{figure}[h]
\includegraphics[width=8.5cm]{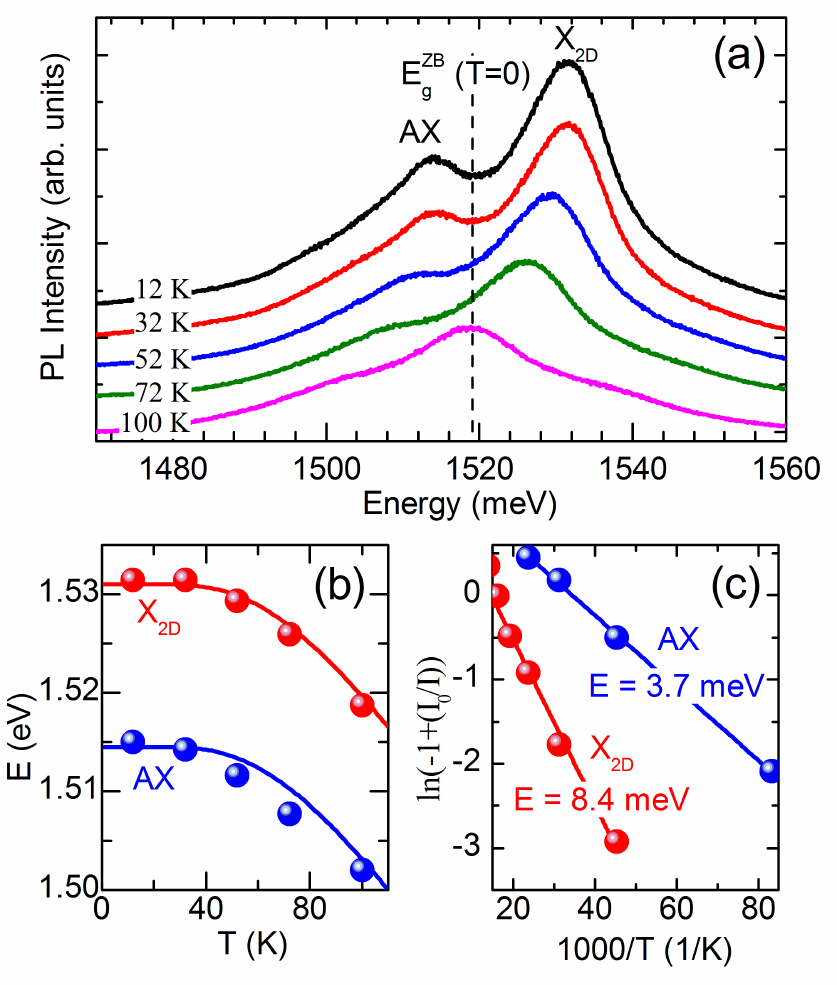}
\caption{\label{fig5}(a) Evolution of PL spectra of core-multi-shell GaAs/AlAs NW as a function of temperature, (b)
Temperature dependence of the emission energy (symbols) together with the established temperature dependence of the
GaAs band gap (lines) (c) Integrated PL intensity rate $\ln(-1 + I_0/I)$ as function of $1/T$ (symbols). The lines are
the linear fits used to deduce the activation energies.}
\end{figure}

The temperature dependence of the $\mu$PL spectra presented in Fig.~\ref{fig5}~(a) provides further confirmation of the
2D character of the observed emission. In the $\mu$PL spectra at T=12~K, two relatively broad peaks are observed
corresponding to the recombination of the acceptor bound exciton (AX) and the confined 2D exciton ($X_{2D}$) at higher
energies. Both shift to lower energy with increasing temperature at a rate which tracks the temperature dependence of
the GaAs band gap (see Fig.~\ref{fig5}~(b)). The emission intensity of AX, observed below $E^{ZB}_{g}$, decreases much
more rapidly than the intensity of the confined exciton peak indicative of a distinctly different dissociation channel
for each transition. The thermal dissociation of excitons leads to a decrease in the normalized integrated emission
intensity $I/I_0 = 1 + \alpha e^{-E/kT}$.~\cite{Bimberg71} In order to estimate the activation energies associated with
both processes, in Fig.~\ref{fig5}~(b) we plot the integrated intensity rate $\ln(-1 + I_0/I)$ as function of $1/T$.
The slope of the linear fit to the data gives the activation energies $E(AX)=3.7$~meV and $E(X_{2D})=8.4$~meV. The
activation energy for AX coincides with the binding energy of the acceptor bound exciton in high purity
GaAs~\cite{Bimberg71} and is also very close to the value reported earlier for ZB GaAs NWs (4~meV).~\cite{Titova06} The
twofold higher activation energy of $X_{2D}$ is expected for 2D confined excitons further confirming the localization
of holes on the facets.

\begin{figure}[h]
\includegraphics[width=11cm]{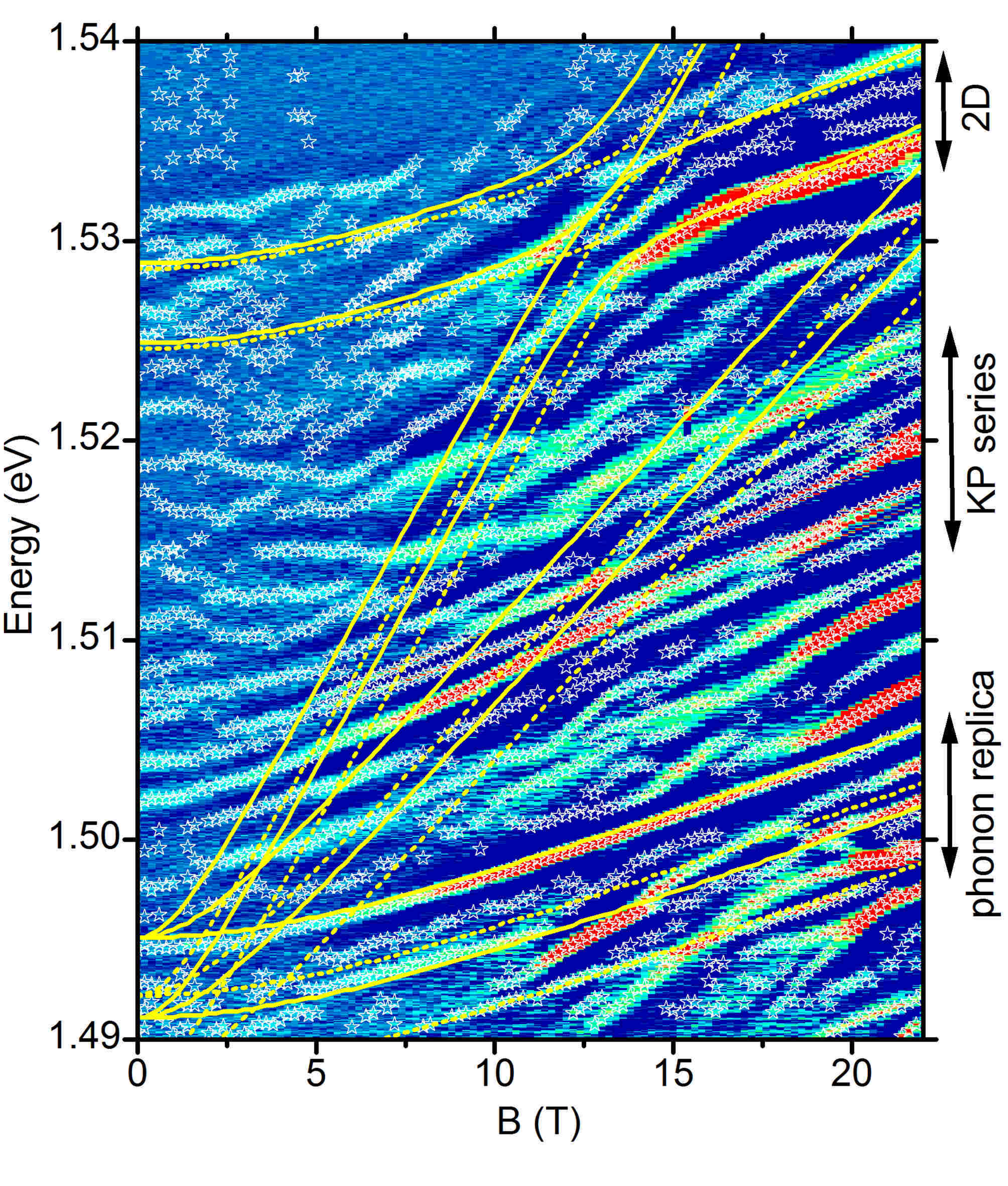}
\caption{\label{fig6}Color plot showing differential $\mu$PL spectra of core-multi-shell GaAs/AlAs NW measured as
function of magnetic field. The excitation power was a few nW and the temperature of the measurements was 1.7~K. The
lines show the calculated evolution of the two high energy 2D emission lines together with their LO (dashed lines) and
TO (solid lines) phonon replicas as described in the text. The observed avoided crossing is the result of resonant
polaron coupling.}
\end{figure}

In Fig.~\ref{fig6} the color plot shows differential $\mu$PL spectra, obtained by subtracting suitably averaged
spectra, in magnetic fields up to $22$~T applied perpendicular to the core-multi-shell NW growth axis. The PL emission
lines sharpen and greatly increase in strength in a magnetic field, which allows us to resolve many more features. On
top of the color plot the symbols (white stars) show the position of peaks manually identified for each spectra. This
is useful to show weak features, particularly at low magnetic fields. At high energies we predominantly observe two
lines corresponding to emission from quantum confined 2D carriers. At intermediate energies there is a series of lines
previously identified with the KP series related to excitons bound to closely spaced defect pairs.~\cite{Plochocka13}
At low energies, as we will later show, the observed lines can be identified with the LO and TO phonon replicas of the
2D emission observed at high energies.

For all the lines, at low magnetic field, the emission energy increases quadratically due to the diamagnetic shift. At
higher magnetic field the dependence becomes linear in the Landau quantization regime. The magnetic field dependence of
the KP series of lines in an undoped NW was discussed in detail in a previous publication.~\cite{Plochocka13} Here we
focus on the lines linked to confined carrier emission. The energy of emission as a function of magnetic field can be
described by the ground state of a 2D harmonic oscillator in perpendicular magnetic field,

\begin{equation}\label{harmonicoscillator}
     E(B) = E_0 + \hbar\sqrt{\omega_{0}^{2}+(\omega_{c}/2)^{2}}
\end{equation}

\noindent where $\omega_{0}$ is the harmonic trap frequency, which
controls the diamagnetic shift and $\omega_{c} = eB/m^{\ast}$ is the
cyclotron frequency. Here we neglect the Zeeman splitting which is
not resolved in our data. The LO and TO phonon replicas are given by
$E(B) - \hbar \omega_{ph}$ where $\hbar \omega_{ph}$ is the LO or TO
phonon energy in GaAs with $\hbar \omega_{LO} = 36.25$~meV or $\hbar
\omega_{TO} = 33.29$~meV.~\cite{Adachi1985} The 2D emission lines
are quite weak at low magnetic field and the lower line shows
distinct evidence for an avoided crossing around $12$~T which makes
it difficult to fit Eq.(1) to the data. Fortunately, the phonon
replicas have reasonable strength over a wide range of magnetic
field allowing us to reliably extract $\hbar \omega_0 = 4.75$~meV
and $m^* = 0.087 m_e$. The second term in the Taylor expansion of
Eq.(1) gives the coefficient for the diamagnetic shift $\hbar e^2 /8
\omega_0 m^{*2} \simeq 50 \mu$eV/T$^2$ which is reasonable for a
confined exciton.~\cite{Walck1998}

Avoided crossings in the emission from the lowest ($n=0$) Landau level have previously been observed in high density
2DEGs and are due to a resonant polaron coupling which occurs when
$\Delta{n}\hbar\omega_{c}=\hbar\omega_{LO}$.~\cite{Harper73} Occupancy arguments require that in order to observe such
an effect the Fermi energy should be similar to or greater than the phonon energy, which requires a high 2D carrier
density $\geq 1 \times 10^{12}$~cm$^{-2}$. The avoided crossing behavior can be described using a perturbation approach
where the unperturbed energies are replaced by

\begin{equation}\label{resonant polaron}
     E_{\pm} = \frac{1}{2}\left ( E + E_{ph}^n \right ) \pm \frac{1}{2} \sqrt{( E - E_{ph}^n)^2 - 4\gamma_{ph}^2}
\end{equation}

\noindent where $E_{ph}^n$ is the energy of the $n^{th}$ phonon
replica obtained from Eq.(1) with $\omega_{c}/2 \rightarrow
(n+\frac{1}{2})\omega_{c}$ and $\gamma_{ph}$ is the characteristic
interaction energy for each phonon. The evolution of the emission
lines, calculated using Eq.(2) and shown by the yellow lines in
Fig.~\ref{fig6}, is in good agreement with our data, nicely
reproducing the phonon replicas and the avoided crossing behavior
observed in the $n=0$ Landau level. Note that as $m^*$ and
$\omega_0$ were extracted from the unperturbed $n=0$ phonon replica
the only fitting parameters are the characteristic interaction
energies ($\gamma_{ph}$) which we find to be equal to $\gamma_{TO}
\simeq 4$~meV and $\gamma_{LO} \simeq 2$~meV. The avoided crossing
with the $n=1$ phonon replica exceeds our maximum magnetic field
occurring for fields above $22$~T. The observed avoided crossing
occurs between the $n=0$ Landau level emission and the $n=2$ Landau
level phonon replica. Moreover, the magnetic field position of the
avoided crossing suggests that interaction with the TO phonon
dominates. The interaction with the LO phonon would lead to an
avoided crossing at higher magnetic field. This is further evidence
for the 2D nature of the carriers involved since coupling to the TO
phonon mode is absent in 3D systems.~\cite{Butov92}

In conclusion, we have carried out optical studies of the single GaAs/AlAs core-multi-shell NW in high magnetic field.
We have compared the obtained results with the typical spectra collected for single GaAs/AlAs core-shell NW with
comparable diameter. In the PL spectra of the core-multi-shell NW we have observed emission above $E^{ZB}_{g}=1.519$~eV
which is related to 2D confinement of carriers in the core. Our results are in good agreement with theoretical
calculations, which predict different localization regimes for carriers as a function of doping. Moreover, in
magneto-PL spectra from core-multi-shell NWs we have observed avoided crossings of emission lines. The underlying
coupling is caused by the hole-phonon interaction in a 2D system with a dense gas. This shows that the presence of 1~nm
GaAs layer in the shell, which acts as an efficient impurity trap, can lead to the efficient incorporation of residual
acceptors (carbon) and the formation of a dense 2D hole gas at the facets of the NW core.

\acknowledgement

We thank Ronit Popovitz-Biro for the careful TEM mesurements. This work was partially supported by the Region
Midi-Pyren\'ees, the Programme Investissements d'Avenir under the program ANR-11-IDEX-0002-02, reference
ANR-10-LABX-0037-NEXT, ANR JCJC project milliPICS and project APOSTD/2013/052 Generalitat Valenciana Vali+d Grant. G.G.
and A.B. acknowledge support from EU-FP7 Initial Training Network INDEX. G.G. acknowledges support from University of
Modena and Reggio emilia, through grant 'Nano- and emerging materials and systems for sustainable technologies'. We
would also like to acknowledge partial support by the Israeli Science Foundation grant \#530/08 and Israeli Ministry of
Science grant \#3-6799.

\suppinfo

\subsection{Experimental techniques}

The study of the optical properties of core-multi-shell NW's was carried out in two experimental setups. For
measurements of the micro-photoluminescence ($\mu$PL) in magnetic field the sample was placed in a system composed of
piezoelectric x-y-z translation stages and a microscope objective. The $\mu$PL system was cooled to a temperature of
T=$1.8$~K in a cryostat placed in a resistive magnet producing magnetic fields of up to $B=22$~T. The field was applied
in the Faraday configuration, perpendicular to the NW $\langle111\rangle$ growth axis. The sample was illuminated by a
diode laser at $660$ nm. Both the exciting and collected light were transmitted through a monomode fiber coupled
directly to the microscope objective. The diameter of the excitation beam on the sample was of the order of 1~$\mu$m.
The emission from the sample was dispersed in a spectrometer equipped with a multichannel CCD camera. For additional
$\mu$PL measurements in the absence of magnetic field, the sample was placed in helium flow cryostat with optical
access. The cryostat was mounted on motorized x-y translation stages. The $\mu$PL was measured for temperatures varying
from $10$ to $100$~K. Excitation and collection was implemented using a microscope objective with a numerical aperture
NA=0.66 and magnification $\times 50$. The diameter of the excitation spot on the sample was of the order of $1 \mu$m
and the $\mu$PL spectra have been recorded using a spectrometer equipped with a multichannel CCD camera.

\subsection{Structural properties of the
GaAs/AlAs/GaAs/AlAs/GaAs core/shell QW nanowires}

\begin{figure}[h]
\includegraphics[width=11cm]{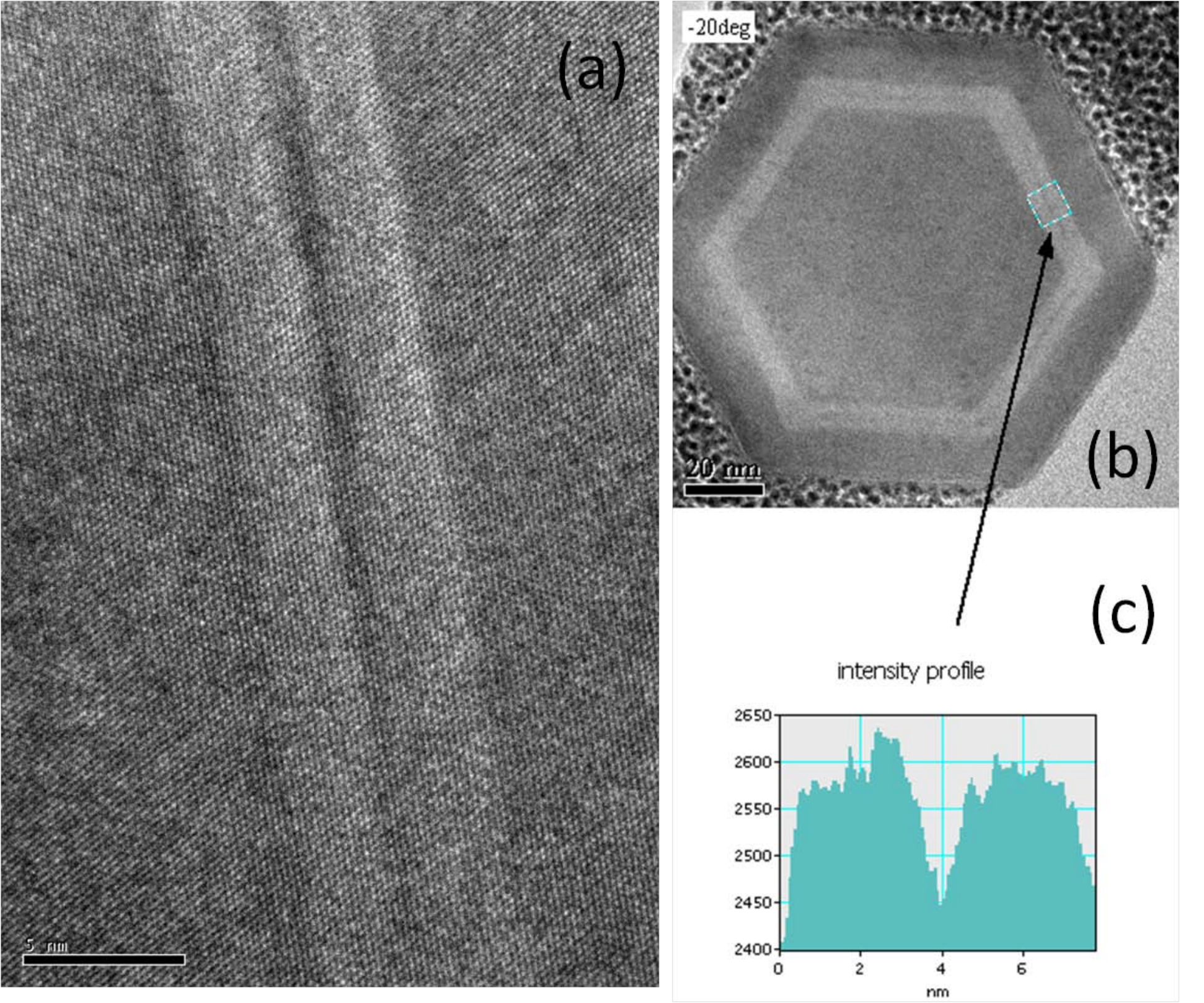}
\caption{(a) HR-TEM image taken on a cross section of the multi
shell nanowire showing the two AlAs shell layers (bright stripes)
and a few monolayers thick GaAs QW embedded between them (GaAs core
and capping layer on right and left hand sides, respectively, scale
bar is 5 nm. (b) TEM image of a cross section taken from a core
multi shell nanowire. (c) intensity profile of the region marked on
the cross section. \label{SuppF1}}
\end{figure}

We do not have a structural characterization of the exact NW
investigated in our $\mu$PL set up. We provide in this section
structural characterization and direct evidence for the presence of
a well-defined QW within a very similar core multi shell structure
grown in the same system and under similar growth conditions and
layer thicknesses. This sample was studied intensively by cross
sectional HR-TEM (prepared using FIB) as demonstrate in
Figure~\ref{SuppF1}. Figure~\ref{SuppF1}(a) shows a HR-TEM image
taken on a cross section of the multi shell nanowire showing the two
AlAs shell layers (bright stripes) and a few monolayers thick GaAs
QW embedded between them (GaAs core and capping layer on the right
and left hand sides, respectively, the scale bar is 5 nm.
Figure~\ref{SuppF1}(b) is a full TEM image of the cross section
taken from a core multi shell nanowire with very similar nominal
thicknesses as the ones studied in this work. Figure~\ref{SuppF1}(c)
shows the intensity profile showing clearly the splitting of the
AlAs layer into two layers, consistent with the presence of a thin
GaAs QW in between them.

\subsection{Self consistent calculation of the charge distribution}

Calculations have been conducted within a standard envelope function approach, in a single-band approximation. We
consider the NW as a 3D system spatially invariant along the NW axis direction $z$. This allows us to factorize the
electron (hole) envelope functions as $\Psi^{e(h)}_{nk_{z}}(\mathbf{r},z)=\varphi^{e(h)}_n(\mathbf{r})e^{ik_{z}z}$,
with parabolic energy dispersions, $E^{e(h)}_{nk_{z}}=\varepsilon^{e(h)}_{n}
\pm\frac{\hbar^{2}k^{2}_{z}}{2m^{z}_{e(h)}}$, in the in-wire momentum $k_{z}$. Over the $\mathbf{r}\equiv x,y$ plane,
the NW cross section is hexagonal, and the material and doping modulations are described in a corresponding hexagonal
domain using a symmetry-compliant triangular grid. We use an isotropic electron effective mass
($m_e^z=m_e^{\mathbf{r}}$) but a highly anisotropic hole mass. Over the in-plane direction we use the hole mass along
the [110] direction which is much larger than the hole mass that we use along the in-wire direction [111].  The ground
state hole density $n_{h}(r)$ is obtained through a Kohn-Sham LDA procedure.

The self-consistent potential $V_{KS}(r)= V(r)+V_{H}(r)+V_{XC}(r)$ includes the effect of the spatial confinement
$V(r)$ determined by the materials band offset, the Hartree potential $V_{H}(r)$ generated by free holes and static
acceptors, and an approximate exchange-correlation potential $V_{XC}$(r).~\cite{Gunnarsson76} In our samples,
conduction band electrons are generated by the laser pumping alone, thus they are minority carriers with a low density:
we solve the corresponding Schr\"odinger equation including the electrostatic potential generated by the
self-consistent hole density and the density of static acceptors. In this way we take into account excitonic effects at
a mean-field level. Further details can be found in references.~\cite{Bertoni11, Royo13}

From the conduction and valence band states, we compute the PL spectra neglecting dynamic screening effects and
assuming that the photoexcited carriers relax to the lowest available state before the radiative recombination. This
means that electrons recombine from the lowest-lying conduction states with holes in states lying above the Fermi level
$\mu$. The PL intensity is obtained as

\begin{equation}
\tau(\omega)\varpropto\sum_{im}|S_{im}|^{2}\int\frac{dk_{z}}{2\pi}
f^{e}(E_{ik_{z}}^{e},T)(1-f^{h}(E_{mk_{z}}^{h}-\mu,T)) \Gamma(E_{ik_{z}}^{e}-E_{mk_{z}}^{h}-\hbar\omega-\gamma),
\label{eq_PL}
\end{equation}

\noindent where

\begin{equation}
S_{im}=\int d\mathbf{r}\varphi_{i}^{e}(\mathbf{r})\varphi_{m}^{h}(\mathbf{r}) \label{eq_overlap}
\end{equation}

\noindent is the overlap integral between a conduction band state $i$ and a valence band state $m$, $f^{e(h)}$ is the
electron (hole) Fermi occupation function at a given temperature T, and $\Gamma$ is a Lorentzian function with a
phenomenological bandwidth $\gamma $ that we set to $1$~meV in order to reproduce the width at half maximum of the
experimental peaks. We use the material-dependent parameters indicated in Table~\ref{table_param}, a temperature
$T=1.8$~K, and we assume that the $1$~nm GaAs quantum well is uniformly doped with a constant density of completely
ionized acceptors $\rho_A$. The Fermi level is placed 0.4~eV above the GaAs valence band edge following experimental
observations in highly p-doped GaAs.~\cite{Pashley93}

\begin{table}[h]
\begin{tabular}{rcc}
  \hline
        & GaAs & AlAs \\
  \hline
  \hline
  Band gap $E_g$ [eV] & 1.519 & 3.02 \\
  Conduction band edge $E_C$ [eV] & 0.9114 & 1.812 \\
  Valence band edge $E_V$[eV] & -0.6076 & -1.208 \\
  Electron effective masses $m^e_z$ & 0.0662 & 0.19 \\
                          $m^e_r$ & 0.0662 & 0.19 \\
  Hole effective masses $m^h_z$ & 0.082 & 0.109 \\
                      $m^h_r$ & 0.680 & 0.818 \\
  \hline
\end{tabular}
\caption{Material parameters employed in the numerical calculations extracted from reference \cite{Levinshtein96}. Band
edges have been calculated assuming the (60:40) rule. Effective masses are taken along [111] direction for $m^{e,h}_z$
and [110] for $m^{e,h}_r$.~\cite{Fishman95}} \label{table_param}
\end{table}


\begin{mcitethebibliography}{54}
\providecommand*\natexlab[1]{#1} \providecommand*\mciteSetBstSublistMode[1]{}
\providecommand*\mciteSetBstMaxWidthForm[2]{} \providecommand*\mciteBstWouldAddEndPuncttrue
  {\def\EndOfBibitem{\unskip.}}
\providecommand*\mciteBstWouldAddEndPunctfalse
  {\let\EndOfBibitem\relax}
\providecommand*\mciteSetBstMidEndSepPunct[3]{} \providecommand*\mciteSetBstSublistLabelBeginEnd[3]{}
\providecommand*\EndOfBibitem{} \mciteSetBstSublistMode{f}
\mciteSetBstMaxWidthForm{subitem}{(\alph{mcitesubitemcount})} \mciteSetBstSublistLabelBeginEnd
  {\mcitemaxwidthsubitemform\space}
  {\relax}
  {\relax}

\bibitem[Hu et~al.(1999)Hu, Odom, and Lieber]{Hu99}
Hu,~J.; Odom,~T.~W.; Lieber,~C.~M. \emph{Accounts of Chemical Research}
  \textbf{1999}, \emph{32}, 435--445\relax
\mciteBstWouldAddEndPuncttrue \mciteSetBstMidEndSepPunct{\mcitedefaultmidpunct}
{\mcitedefaultendpunct}{\mcitedefaultseppunct}\relax \EndOfBibitem
\bibitem[Cui and Lieber(2001)Cui, and Lieber]{Cui01}
Cui,~Y.; Lieber,~C.~M. \emph{Science} \textbf{2001}, \emph{291}, 851\relax \mciteBstWouldAddEndPuncttrue
\mciteSetBstMidEndSepPunct{\mcitedefaultmidpunct} {\mcitedefaultendpunct}{\mcitedefaultseppunct}\relax \EndOfBibitem
\bibitem[Thelander(2006)]{Therlander06}
Thelander,~C. \emph{Mater. Today} \textbf{2006}, \emph{9}, 28\relax \mciteBstWouldAddEndPuncttrue
\mciteSetBstMidEndSepPunct{\mcitedefaultmidpunct} {\mcitedefaultendpunct}{\mcitedefaultseppunct}\relax \EndOfBibitem
\bibitem[Lieber and Wang(2007)Lieber, and Wang]{Lieber07}
Lieber,~C.~M.; Wang,~Z.~L. \emph{MRS Bulletin} \textbf{2007}, \emph{32},
  99--108\relax
\mciteBstWouldAddEndPuncttrue \mciteSetBstMidEndSepPunct{\mcitedefaultmidpunct}
{\mcitedefaultendpunct}{\mcitedefaultseppunct}\relax \EndOfBibitem
\bibitem[Lu and Lieber(2007)Lu, and Lieber]{Lu07}
Lu,~W.; Lieber,~C.~M. \emph{Nature Mater.} \textbf{2007}, \emph{6}, 841\relax \mciteBstWouldAddEndPuncttrue
\mciteSetBstMidEndSepPunct{\mcitedefaultmidpunct} {\mcitedefaultendpunct}{\mcitedefaultseppunct}\relax \EndOfBibitem
\bibitem[Tans et~al.(1998)Tans, Verschueren, and Dekker]{Tans98}
Tans,~S.~J.; Verschueren,~A. R.~M.; Dekker,~C. \emph{Nature} \textbf{1998},
  \emph{393}, 49\relax
\mciteBstWouldAddEndPuncttrue \mciteSetBstMidEndSepPunct{\mcitedefaultmidpunct}
{\mcitedefaultendpunct}{\mcitedefaultseppunct}\relax \EndOfBibitem
\bibitem[Cui et~al.(2003)Cui, Zhong, Wang, Wang, and Lieber]{Cui03}
Cui,~Y.; Zhong,~Z.; Wang,~D.; Wang,~W.~U.; Lieber,~C.~M. \emph{Nano Letters}
  \textbf{2003}, \emph{3}, 149--152\relax
\mciteBstWouldAddEndPuncttrue \mciteSetBstMidEndSepPunct{\mcitedefaultmidpunct}
{\mcitedefaultendpunct}{\mcitedefaultseppunct}\relax \EndOfBibitem
\bibitem[Duan et~al.(2001)Duan, Huang, Cui, Wang, and Lieber]{Duan01}
Duan,~X.; Huang,~Y.; Cui,~Y.; Wang,~J.; Lieber,~C.~M. \emph{Nature}
  \textbf{2001}, \emph{409}, 66\relax
\mciteBstWouldAddEndPuncttrue \mciteSetBstMidEndSepPunct{\mcitedefaultmidpunct}
{\mcitedefaultendpunct}{\mcitedefaultseppunct}\relax \EndOfBibitem
\bibitem[Krogstrup et~al.(2013)Krogstrup, J{\o}rgensen, Heiss, Demichel, Holm,
  Aagesen, Nygard, and i~Morral]{Krogstrup13}
Krogstrup,~P.; J{\o}rgensen,~H.~I.; Heiss,~M.; Demichel,~O.; Holm,~J.~V.;
  Aagesen,~M.; Nygard,~J.; i~Morral,~A.~F. \emph{Nature Photonics}
  \textbf{2013}, \emph{7}, 306--310\relax
\mciteBstWouldAddEndPuncttrue \mciteSetBstMidEndSepPunct{\mcitedefaultmidpunct}
{\mcitedefaultendpunct}{\mcitedefaultseppunct}\relax \EndOfBibitem
\bibitem[Spirkoska et~al.(2011)Spirkoska, Fontcuberta~i Morral, Dufouleur, Xie,
  and Abstreiter]{Spirkoska11}
Spirkoska,~D.; Fontcuberta~i Morral,~A.; Dufouleur,~J.; Xie,~Q.; Abstreiter,~G.
  \emph{physica status solidi (RRL) Rapid Research Letters} \textbf{2011},
  \emph{5}, 353--355\relax
\mciteBstWouldAddEndPuncttrue \mciteSetBstMidEndSepPunct{\mcitedefaultmidpunct}
{\mcitedefaultendpunct}{\mcitedefaultseppunct}\relax \EndOfBibitem
\bibitem[Fickenscher et~al.(2013)Fickenscher, Shi, Jackson, Smith,
  Yarrison-Rice, Zheng, Miller, Etheridge, Wong, Gao, Deshpande, Tan, and
  Jagadish]{Fickenscher13}
Fickenscher,~M.; Shi,~T.; Jackson,~H.~E.; Smith,~L.~M.; Yarrison-Rice,~J.~M.;
  Zheng,~C.; Miller,~P.; Etheridge,~J.; Wong,~B.~M.; Gao,~Q.; Deshpande,~S.;
  Tan,~H.~H.; Jagadish,~C. \emph{Nano Letters} \textbf{2013}, \emph{13},
  1016--1022\relax
\mciteBstWouldAddEndPuncttrue \mciteSetBstMidEndSepPunct{\mcitedefaultmidpunct}
{\mcitedefaultendpunct}{\mcitedefaultseppunct}\relax \EndOfBibitem
\bibitem[Estrin et~al.(2013)Estrin, Rich, Kretinin, and Shtrikman]{Estrin13}
Estrin,~Y.; Rich,~D.~H.; Kretinin,~A.~V.; Shtrikman,~H. \emph{Nano Letters}
  \textbf{2013}, \emph{13}, 1602--1610\relax
\mciteBstWouldAddEndPuncttrue \mciteSetBstMidEndSepPunct{\mcitedefaultmidpunct}
{\mcitedefaultendpunct}{\mcitedefaultseppunct}\relax \EndOfBibitem
\bibitem[Fontcuberta~i Morral et~al.(2008)Fontcuberta~i Morral, Spirkoska,
  Arbiol, Heigoldt, Morante, and Abstreiter]{FontcubertaiMorral08}
Fontcuberta~i Morral,~A.; Spirkoska,~D.; Arbiol,~J.; Heigoldt,~M.;
  Morante,~J.~R.; Abstreiter,~G. \emph{Small} \textbf{2008}, \emph{4},
  899--903\relax
\mciteBstWouldAddEndPuncttrue \mciteSetBstMidEndSepPunct{\mcitedefaultmidpunct}
{\mcitedefaultendpunct}{\mcitedefaultseppunct}\relax \EndOfBibitem
\bibitem[Klitzing et~al.(1980)Klitzing, Dorda, and Pepper]{Klitzing1980}
Klitzing,~K.~v.; Dorda,~G.; Pepper,~M. \emph{Phys. Rev. Lett.} \textbf{1980},
  \emph{45}, 494--497\relax
\mciteBstWouldAddEndPuncttrue \mciteSetBstMidEndSepPunct{\mcitedefaultmidpunct}
{\mcitedefaultendpunct}{\mcitedefaultseppunct}\relax \EndOfBibitem
\bibitem[Tsui et~al.(1982)Tsui, Stormer, and Gossard]{Tsui1982}
Tsui,~D.~C.; Stormer,~H.~L.; Gossard,~A.~C. \emph{Phys. Rev. Lett.}
  \textbf{1982}, \emph{48}, 1559--1562\relax
\mciteBstWouldAddEndPuncttrue \mciteSetBstMidEndSepPunct{\mcitedefaultmidpunct}
{\mcitedefaultendpunct}{\mcitedefaultseppunct}\relax \EndOfBibitem
\bibitem[Titova et~al.(2006)Titova, Hoang, Jackson, Smith, Yarrison-Rice, Kim,
  Joyce, Tan, and Jagadish]{Titova06}
Titova,~L.~V.; Hoang,~T.~B.; Jackson,~H.~E.; Smith,~L.~M.;
  Yarrison-Rice,~J.~M.; Kim,~Y.; Joyce,~H.~J.; Tan,~H.~H.; Jagadish,~C.
  \emph{Applied Physics Letters} \textbf{2006}, \emph{89}, --\relax
\mciteBstWouldAddEndPuncttrue \mciteSetBstMidEndSepPunct{\mcitedefaultmidpunct}
{\mcitedefaultendpunct}{\mcitedefaultseppunct}\relax \EndOfBibitem
\bibitem[Pistol and Pryor(2008)Pistol, and Pryor]{Pistol08}
Pistol,~M.-E.; Pryor,~C.~E. \emph{Phys. Rev. B} \textbf{2008}, \emph{78},
  115319\relax
\mciteBstWouldAddEndPuncttrue \mciteSetBstMidEndSepPunct{\mcitedefaultmidpunct}
{\mcitedefaultendpunct}{\mcitedefaultseppunct}\relax \EndOfBibitem
\bibitem[Shtrikman et~al.(2009)Shtrikman, Popovitz-Biro, Kretinin, and
  Heiblum]{Shtrikman09a}
Shtrikman,~H.; Popovitz-Biro,~R.; Kretinin,~A.; Heiblum,~M. \emph{Nano Letters}
  \textbf{2009}, \emph{9}, 215--219\relax
\mciteBstWouldAddEndPuncttrue \mciteSetBstMidEndSepPunct{\mcitedefaultmidpunct}
{\mcitedefaultendpunct}{\mcitedefaultseppunct}\relax \EndOfBibitem
\bibitem[Shtrikman et~al.(2009)Shtrikman, Popovitz-Biro, Kretinin, Houben,
  Heiblum, Bukala, Galicka, Buczko, and Kacman]{Shtrikman09}
Shtrikman,~H.; Popovitz-Biro,~R.; Kretinin,~A.; Houben,~L.; Heiblum,~M.;
  Bukala,~M.; Galicka,~M.; Buczko,~R.; Kacman,~P. \emph{Nano Letters}
  \textbf{2009}, \emph{9}, 1506--1510, PMID: 19253998\relax
\mciteBstWouldAddEndPuncttrue \mciteSetBstMidEndSepPunct{\mcitedefaultmidpunct}
{\mcitedefaultendpunct}{\mcitedefaultseppunct}\relax \EndOfBibitem
\bibitem[Spirkoska et~al.(2009)Spirkoska, Arbiol, Gustafsson, Conesa-Boj, Glas,
  Zardo, Heigoldt, Gass, Bleloch, Estrade, Kaniber, Rossler, Peiro, Morante,
  Abstreiter, Samuelson, and Fontcuberta~i Morral]{Spirkoska09}
Spirkoska,~D. et~al.  \emph{Phys. Rev. B} \textbf{2009}, \emph{80},
  245325\relax
\mciteBstWouldAddEndPuncttrue \mciteSetBstMidEndSepPunct{\mcitedefaultmidpunct}
{\mcitedefaultendpunct}{\mcitedefaultseppunct}\relax \EndOfBibitem
\bibitem[Algra et~al.(2011)Algra, Hocevar, Verheijen, Zardo, Immink, van
  Enckevort, Abstreiter, Kouwenhoven, Vlieg, and Bakkers]{Algra11}
Algra,~R.~E.; Hocevar,~M.; Verheijen,~M.~A.; Zardo,~I.; Immink,~G. G.~W.; van
  Enckevort,~W. J.~P.; Abstreiter,~G.; Kouwenhoven,~L.~P.; Vlieg,~E.;
  Bakkers,~E. P. A.~M. \emph{Nano Letters} \textbf{2011}, \emph{11},
  1690--1694\relax
\mciteBstWouldAddEndPuncttrue \mciteSetBstMidEndSepPunct{\mcitedefaultmidpunct}
{\mcitedefaultendpunct}{\mcitedefaultseppunct}\relax \EndOfBibitem
\bibitem[Jahn et~al.(2012)Jahn, L\"ahnemann, Pf\"uller, Brandt, Breuer,
  Jenichen, Ramsteiner, Geelhaar, and Riechert]{Jahn12}
Jahn,~U.; L\"ahnemann,~J.; Pf\"uller,~C.; Brandt,~O.; Breuer,~S.; Jenichen,~B.;
  Ramsteiner,~M.; Geelhaar,~L.; Riechert,~H. \emph{Phys. Rev. B} \textbf{2012},
  \emph{85}, 045323\relax
\mciteBstWouldAddEndPuncttrue \mciteSetBstMidEndSepPunct{\mcitedefaultmidpunct}
{\mcitedefaultendpunct}{\mcitedefaultseppunct}\relax \EndOfBibitem
\bibitem[Musin and Filler(2012)Musin, and Filler]{Musin12}
Musin,~I.~R.; Filler,~M.~A. \emph{Nano Letters} \textbf{2012}, \emph{12},
  3363--3368\relax
\mciteBstWouldAddEndPuncttrue \mciteSetBstMidEndSepPunct{\mcitedefaultmidpunct}
{\mcitedefaultendpunct}{\mcitedefaultseppunct}\relax \EndOfBibitem
\bibitem[Assali et~al.(2013)Assali, Zardo, Plissard, Kriegner, Verheijen,
  Bauer, Meijerink, Belabbes, Bechstedt, Haverkort, and Bakkers]{Assali13}
Assali,~S.; Zardo,~I.; Plissard,~S.; Kriegner,~D.; Verheijen,~M.~A.; Bauer,~G.;
  Meijerink,~A.; Belabbes,~A.; Bechstedt,~F.; Haverkort,~J. E.~M.; Bakkers,~E.
  P. A.~M. \emph{Nano Letters} \textbf{2013}, \emph{13}, 1559--1563\relax
\mciteBstWouldAddEndPuncttrue \mciteSetBstMidEndSepPunct{\mcitedefaultmidpunct}
{\mcitedefaultendpunct}{\mcitedefaultseppunct}\relax \EndOfBibitem
\bibitem[Hilse et~al.(2010)Hilse, Ramsteiner, Breuer, Geelhaar, and
  Riechert]{Hilse10}
Hilse,~M.; Ramsteiner,~M.; Breuer,~S.; Geelhaar,~L.; Riechert,~H. \emph{Applied
  Physics Letters} \textbf{2010}, \emph{96}, --\relax
\mciteBstWouldAddEndPuncttrue \mciteSetBstMidEndSepPunct{\mcitedefaultmidpunct}
{\mcitedefaultendpunct}{\mcitedefaultseppunct}\relax \EndOfBibitem
\bibitem[Dufouleur et~al.(2010)Dufouleur, Colombo, Garma, Ketterer, Uccelli,
  Nicotra, and Fontcuberta~i Morral]{Dufouleur10}
Dufouleur,~J.; Colombo,~C.; Garma,~T.; Ketterer,~B.; Uccelli,~E.; Nicotra,~M.;
  Fontcuberta~i Morral,~A. \emph{Nano Letters} \textbf{2010}, \emph{10},
  1734--1740\relax
\mciteBstWouldAddEndPuncttrue \mciteSetBstMidEndSepPunct{\mcitedefaultmidpunct}
{\mcitedefaultendpunct}{\mcitedefaultseppunct}\relax \EndOfBibitem
\bibitem[Casadei et~al.(2013)Casadei, Krogstrup, Heiss, R{\"o}hr, Colombo,
  Ruelle, Upadhyay, S{\o}rensen, Nyg{\aa}rd, and Fontcuberta~i
  Morral]{Casadei13}
Casadei,~A.; Krogstrup,~P.; Heiss,~M.; R{\"o}hr,~J.~A.; Colombo,~C.;
  Ruelle,~T.; Upadhyay,~S.; S{\o}rensen,~C.~B.; Nyg{\aa}rd,~J.; Fontcuberta~i
  Morral,~A. \emph{Applied Physics Letters} \textbf{2013}, \emph{102}, --\relax
\mciteBstWouldAddEndPuncttrue \mciteSetBstMidEndSepPunct{\mcitedefaultmidpunct}
{\mcitedefaultendpunct}{\mcitedefaultseppunct}\relax \EndOfBibitem
\bibitem[Heiss et~al.(2013)Heiss, Fontana, Gustafsson, W\"ust, Magen, O'Regan,
  Luo, Ketterer, Conesa-Boj, Kuhlmann, Houel, Russo-Averchi, Morante, Cantoni,
  Marzari, Arbiol, Zunger, Warburton, and i~Morral]{Heiss13}
Heiss,~M. et~al.  \emph{Nature Materials} \textbf{2013}, \emph{12}, 439\relax \mciteBstWouldAddEndPuncttrue
\mciteSetBstMidEndSepPunct{\mcitedefaultmidpunct} {\mcitedefaultendpunct}{\mcitedefaultseppunct}\relax \EndOfBibitem
\bibitem[Rudolph et~al.(2013)Rudolph, Funk, Döblinger, Morkötter, Hertenberger,
  Schweickert, Becker, Matich, Bichler, Spirkoska, Zardo, Finley, Abstreiter,
  and Koblmüller]{Funk13}
Rudolph,~D.; Funk,~S.; Döblinger,~M.; Morkötter,~S.; Hertenberger,~S.;
  Schweickert,~L.; Becker,~J.; Matich,~S.; Bichler,~M.; Spirkoska,~D.;
  Zardo,~I.; Finley,~J.~J.; Abstreiter,~G.; Koblmüller,~G. \emph{Nano Letters}
  \textbf{2013}, \emph{13}, 1522--1527\relax
\mciteBstWouldAddEndPuncttrue \mciteSetBstMidEndSepPunct{\mcitedefaultmidpunct}
{\mcitedefaultendpunct}{\mcitedefaultseppunct}\relax \EndOfBibitem
\bibitem[Hocevar et~al.(2013)Hocevar, Thanh~Giang, Songmuang, den Hertog,
  Besombes, Bleuse, Niquet, and Pelekanos]{Hocevar12}
Hocevar,~M.; Thanh~Giang,~L.~T.; Songmuang,~R.; den Hertog,~M.; Besombes,~L.;
  Bleuse,~J.; Niquet,~Y.-M.; Pelekanos,~N.~T. \emph{Applied Physics Letters}
  \textbf{2013}, \emph{102}, --\relax
\mciteBstWouldAddEndPuncttrue \mciteSetBstMidEndSepPunct{\mcitedefaultmidpunct}
{\mcitedefaultendpunct}{\mcitedefaultseppunct}\relax \EndOfBibitem
\bibitem[Funk et~al.(2013)Funk, Royo, Zardo, Rudolph, Mork\"otter, Mayer,
  Becker, Bechtold, Matich, D\"oblinger, Bichler, Koblm\"uller, Finley,
  Bertoni, Goldoni, and Abstreiter]{Funk2013}
Funk,~S. et~al.  \emph{Nano Letters} \textbf{2013}, \emph{13}, 6189--6196\relax \mciteBstWouldAddEndPuncttrue
\mciteSetBstMidEndSepPunct{\mcitedefaultmidpunct} {\mcitedefaultendpunct}{\mcitedefaultseppunct}\relax \EndOfBibitem
\bibitem[Plochocka et~al.(2013)Plochocka, Mitioglu, Maude, Rikken, Granados~del
  Aguila, Christianen, Kacman, and Shtrikman]{Plochocka13}
Plochocka,~P.; Mitioglu,~A.~A.; Maude,~D.~K.; Rikken,~G. L. J.~A.; Granados~del
  Aguila,~A.; Christianen,~P. C.~M.; Kacman,~P.; Shtrikman,~H. \emph{Nano
  Letters} \textbf{2013}, \emph{13}, 2442--2447\relax
\mciteBstWouldAddEndPuncttrue \mciteSetBstMidEndSepPunct{\mcitedefaultmidpunct}
{\mcitedefaultendpunct}{\mcitedefaultseppunct}\relax \EndOfBibitem
\bibitem[Kunzel and Ploog(1980)Kunzel, and Ploog]{Kunzel80}
Kunzel,~H.; Ploog,~K. \emph{Applied Physics Letters} \textbf{1980}, \emph{37},
  416--418\relax
\mciteBstWouldAddEndPuncttrue \mciteSetBstMidEndSepPunct{\mcitedefaultmidpunct}
{\mcitedefaultendpunct}{\mcitedefaultseppunct}\relax \EndOfBibitem
\bibitem[Skolnick et~al.(1985)Skolnick, Harris, Tu, Brennan, and
  Sturge]{Skolnick85}
Skolnick,~M.~S.; Harris,~T.~D.; Tu,~C.~W.; Brennan,~T.~M.; Sturge,~M.~D.
  \emph{Applied Physics Letters} \textbf{1985}, \emph{46}, 427--429\relax
\mciteBstWouldAddEndPuncttrue \mciteSetBstMidEndSepPunct{\mcitedefaultmidpunct}
{\mcitedefaultendpunct}{\mcitedefaultseppunct}\relax \EndOfBibitem
\bibitem[Pavesi and Guzzi(1994)Pavesi, and Guzzi]{Pavesi94}
Pavesi,~L.; Guzzi,~M. \emph{Journal of Applied Physics} \textbf{1994},
  \emph{75}, 4779--4842\relax
\mciteBstWouldAddEndPuncttrue \mciteSetBstMidEndSepPunct{\mcitedefaultmidpunct}
{\mcitedefaultendpunct}{\mcitedefaultseppunct}\relax \EndOfBibitem
\bibitem[Sager et~al.(2013)Sager, Gutsche, Prost, Tegude, and Bacher]{Sager13}
Sager,~D.; Gutsche,~C.; Prost,~W.; Tegude,~F.-J.; Bacher,~G. \emph{Journal of
  Applied Physics} \textbf{2013}, \emph{113}, --\relax
\mciteBstWouldAddEndPuncttrue \mciteSetBstMidEndSepPunct{\mcitedefaultmidpunct}
{\mcitedefaultendpunct}{\mcitedefaultseppunct}\relax \EndOfBibitem
\bibitem[Borghs et~al.(1989)Borghs, Bhattacharyya, Deneffe, Van~Mieghem, and
  Mertens]{Borghs89}
Borghs,~G.; Bhattacharyya,~K.; Deneffe,~K.; Van~Mieghem,~P.; Mertens,~R.
  \emph{Journal of Applied Physics} \textbf{1989}, \emph{66}, 4381--4386\relax
\mciteBstWouldAddEndPuncttrue \mciteSetBstMidEndSepPunct{\mcitedefaultmidpunct}
{\mcitedefaultendpunct}{\mcitedefaultseppunct}\relax \EndOfBibitem
\bibitem[Ferreira et~al.(1996)Ferreira, Holtz, Sernelius, Buyanova, Monemar,
  Mauritz, Ekenberg, Sundaram, Campman, Merz, and Gossard]{Ferreira1996}
Ferreira,~A.~C.; Holtz,~P.~O.; Sernelius,~B.~E.; Buyanova,~I.; Monemar,~B.;
  Mauritz,~O.; Ekenberg,~U.; Sundaram,~M.; Campman,~K.; Merz,~J.~L.;
  Gossard,~A.~C. \emph{Phys. Rev. B} \textbf{1996}, \emph{54},
  16989--16993\relax
\mciteBstWouldAddEndPuncttrue \mciteSetBstMidEndSepPunct{\mcitedefaultmidpunct}
{\mcitedefaultendpunct}{\mcitedefaultseppunct}\relax \EndOfBibitem
\bibitem[Glasberg et~al.(2001)Glasberg, Shtrikman, and
  Bar-Joseph]{Glasberg2001}
Glasberg,~S.; Shtrikman,~H.; Bar-Joseph,~I. \emph{Phys. Rev. B} \textbf{2001},
  \emph{63}, 201308\relax
\mciteBstWouldAddEndPuncttrue \mciteSetBstMidEndSepPunct{\mcitedefaultmidpunct}
{\mcitedefaultendpunct}{\mcitedefaultseppunct}\relax \EndOfBibitem
\bibitem[Meynadier et~al.(1985)Meynadier, Brum, Delalande, Voos, Alexandre, and
  Li\'{e}vin]{Meynadier85}
Meynadier,~M.~H.; Brum,~J.~A.; Delalande,~C.; Voos,~M.; Alexandre,~F.;
  Li\'{e}vin,~J.~L. \emph{Journal of Applied Physics} \textbf{1985}, \emph{58},
  4307--4312\relax
\mciteBstWouldAddEndPuncttrue \mciteSetBstMidEndSepPunct{\mcitedefaultmidpunct}
{\mcitedefaultendpunct}{\mcitedefaultseppunct}\relax \EndOfBibitem
\bibitem[Petroff et~al.(1984)Petroff, Miller, Gossard, and Wiegmann]{Petroff84}
Petroff,~P.~M.; Miller,~R.~C.; Gossard,~A.~C.; Wiegmann,~W. \emph{Applied
  Physics Letters} \textbf{1984}, \emph{44}, 217--219\relax
\mciteBstWouldAddEndPuncttrue \mciteSetBstMidEndSepPunct{\mcitedefaultmidpunct}
{\mcitedefaultendpunct}{\mcitedefaultseppunct}\relax \EndOfBibitem
\bibitem[Tokumitsu et~al.(1986)Tokumitsu, Katoh, Kimura, Konagai, and
  Takahashi]{Tokumitsu86}
Tokumitsu,~E.; Katoh,~T.; Kimura,~R.; Konagai,~M.; Takahashi,~K. \emph{Japanese
  Journal of Applied Physics} \textbf{1986}, \emph{25}, 1211--1215\relax
\mciteBstWouldAddEndPuncttrue \mciteSetBstMidEndSepPunct{\mcitedefaultmidpunct}
{\mcitedefaultendpunct}{\mcitedefaultseppunct}\relax \EndOfBibitem
\bibitem[Bertoni et~al.(2011)Bertoni, Royo, Mahawish, and Goldoni]{Bertoni11}
Bertoni,~A.; Royo,~M.; Mahawish,~F.; Goldoni,~G. \emph{Phys. Rev. B}
  \textbf{2011}, \emph{84}, 205323\relax
\mciteBstWouldAddEndPuncttrue \mciteSetBstMidEndSepPunct{\mcitedefaultmidpunct}
{\mcitedefaultendpunct}{\mcitedefaultseppunct}\relax \EndOfBibitem
\bibitem[Bimberg et~al.(1971)Bimberg, Sondergeld, and Grobe]{Bimberg71}
Bimberg,~D.; Sondergeld,~M.; Grobe,~E. \emph{Phys. Rev. B} \textbf{1971},
  \emph{4}, 3451--3455\relax
\mciteBstWouldAddEndPuncttrue \mciteSetBstMidEndSepPunct{\mcitedefaultmidpunct}
{\mcitedefaultendpunct}{\mcitedefaultseppunct}\relax \EndOfBibitem
\bibitem[Adachi(1985)]{Adachi1985}
Adachi,~S. \emph{Journal of Applied Physics} \textbf{1985}, \emph{58},
  R1--R29\relax
\mciteBstWouldAddEndPuncttrue \mciteSetBstMidEndSepPunct{\mcitedefaultmidpunct}
{\mcitedefaultendpunct}{\mcitedefaultseppunct}\relax \EndOfBibitem
\bibitem[Walck and Reinecke(1998)Walck, and Reinecke]{Walck1998}
Walck,~S.~N.; Reinecke,~T.~L. \emph{Phys. Rev. B} \textbf{1998}, \emph{57},
  9088--9096\relax
\mciteBstWouldAddEndPuncttrue \mciteSetBstMidEndSepPunct{\mcitedefaultmidpunct}
{\mcitedefaultendpunct}{\mcitedefaultseppunct}\relax \EndOfBibitem
\bibitem[Harper et~al.(1973)Harper, Hodby, and Stradling]{Harper73}
Harper,~P.~G.; Hodby,~J.~W.; Stradling,~R.~A. \emph{Reports on Progress in
  Physics} \textbf{1973}, \emph{36}, 1\relax
\mciteBstWouldAddEndPuncttrue \mciteSetBstMidEndSepPunct{\mcitedefaultmidpunct}
{\mcitedefaultendpunct}{\mcitedefaultseppunct}\relax \EndOfBibitem
\bibitem[Butov et~al.(1992)Butov, Grinev, Kulakovskii, and Andersson]{Butov92}
Butov,~L.~V.; Grinev,~V.~I.; Kulakovskii,~V.~D.; Andersson,~T.~G. \emph{Phys.
  Rev. B} \textbf{1992}, \emph{46}, 13627--13630\relax
\mciteBstWouldAddEndPuncttrue \mciteSetBstMidEndSepPunct{\mcitedefaultmidpunct}
{\mcitedefaultendpunct}{\mcitedefaultseppunct}\relax \EndOfBibitem
\bibitem[Gunnarsson and Lundqvist(1976)Gunnarsson, and Lundqvist]{Gunnarsson76}
Gunnarsson,~O.; Lundqvist,~B.~I. \emph{Phys. Rev. B} \textbf{1976}, \emph{13},
  4274--4298\relax
\mciteBstWouldAddEndPuncttrue \mciteSetBstMidEndSepPunct{\mcitedefaultmidpunct}
{\mcitedefaultendpunct}{\mcitedefaultseppunct}\relax \EndOfBibitem
\bibitem[Royo et~al.(2013)Royo, Bertoni, and Goldoni]{Royo13}
Royo,~M.; Bertoni,~A.; Goldoni,~G. \emph{Phys. Rev. B} \textbf{2013},
  \emph{87}, 115316\relax
\mciteBstWouldAddEndPuncttrue \mciteSetBstMidEndSepPunct{\mcitedefaultmidpunct}
{\mcitedefaultendpunct}{\mcitedefaultseppunct}\relax \EndOfBibitem
\bibitem[Pashley et~al.(1993)Pashley, Haberern, Feenstra, and
  Kirchner]{Pashley93}
Pashley,~M.~D.; Haberern,~K.~W.; Feenstra,~R.~M.; Kirchner,~P.~D. \emph{Phys.
  Rev. B} \textbf{1993}, \emph{48}, 4612--4615\relax
\mciteBstWouldAddEndPuncttrue \mciteSetBstMidEndSepPunct{\mcitedefaultmidpunct}
{\mcitedefaultendpunct}{\mcitedefaultseppunct}\relax \EndOfBibitem
\bibitem[Levinshtein et~al.(1996)Levinshtein, Rumyantsev, and
  Shur]{Levinshtein96}
Levinshtein,~M.; Rumyantsev,~S.; Shur,~M. \emph{Handbook series on
  semiconductor parameters Vol. 2}; World Scientific Publishing, 1996\relax
\mciteBstWouldAddEndPuncttrue \mciteSetBstMidEndSepPunct{\mcitedefaultmidpunct}
{\mcitedefaultendpunct}{\mcitedefaultseppunct}\relax \EndOfBibitem
\bibitem[Fishman(1995)]{Fishman95}
Fishman,~G. \emph{Phys. Rev. B} \textbf{1995}, \emph{52}, 11132--11143\relax \mciteBstWouldAddEndPuncttrue
\mciteSetBstMidEndSepPunct{\mcitedefaultmidpunct} {\mcitedefaultendpunct}{\mcitedefaultseppunct}\relax \EndOfBibitem
\end{mcitethebibliography}

\providecommand*\mcitethebibliography{\thebibliography} \csname @ifundefined\endcsname{endmcitethebibliography}
  {\let\endmcitethebibliography\endthebibliography}{}

\end{document}